\documentclass[12pt]{article}
\usepackage{graphicx}
\begin{document}

{\bf Applications of Physics and Mathematics to Social Science}

\bigskip

D. Stauffer* and S. Solomon

\bigskip
Racah Institute of Physics, Hebrew University, IL-91904 Jerusalem, Israel

\bigskip
*Institute for Theoretical Physics, Cologne University, 

D-50923 K\"oln, Euroland

\bigskip
{\bf \large Article Outline}
\bigskip

Glossary

\bigskip
I. Definition 

II. Introduction

III. Some models and concepts

IV. Applications

V. Future Directions

\bigskip
{\bf \large Glossary}
\bigskip

{\bf Cellular Automata}

Discrete variables on a discrete lattice change in discrete time steps.

\medskip

{\bf Ising model} 

Neighbouring variables prefer to be the same but exceptions are possible. The
probability for such exceptions is an exponential function of "temperature".

\medskip
{\bf Percolation} 

Each site is randomly either occupied or empty, leading to random clusters. At 
the percolation threshold for the first time an infinite cluster is formed.
\medskip

{\bf Universality} 

Certain properties are the same for a whole set of models or of real objects.

\bigskip
\section{Definition}

This article introduces into the whole section on Social Sciences, edited by
A. Nowak for this Encyclopedia, concentrating on the applications of 
mathematics and physics. Here under "mathematics" we include also all computer simulations if they are not taken from physics, while physics applications
include simulations of models which basically existed already in physics before
they were applied to social simulations. Thus obviously there is no sharp 
border between applications from physics and from mathematics in the sense of 
our definition. Also social science is not defined precisely. We will include 
some economics as well as some linguistics, but not social insects or fish 
swarms, nor human epidemics or demography. Also, we mention not only this 
section by also the section on agent-based modelling edited by F. Castiglione 
as containing articles of social interest.

\section{Introduction}

If mathematical/physical methods are applied to social sciences, a major 
problem is the mutual lack of literature knowledge. Take for example the 
Schelling model of racial segregation in cities \cite{schelling}. Sociologist
don't cite the better and simpler Ising model, physicists ignored the Schelling
model for decades, and sociologists also ignored better sociology work 
\cite{jones}. For simulations of 
financial markets, many econophysicists thought that they had introduced 
Monte Carlo and agent-based simulations to finance, not knowing of earlier 
work from some forward-looking Nobel laureates in economics 
\cite{stigler,markowitz}. For inter-community relations, already 25 centuries
ago analogies with liquids were pointed out by Empedokles in Sicily
\cite{stauffer}. More recently, Ettore Majorana \cite{mantegna} around 1940
suggested to apply quantum-mechanical uncertainty to socio-economic questions.
With emphasis shifted to statistical physics, sociophysics and econophysics
became fashionable around the change of the millennium, but continuous lines
of research by some physicists started \cite{weidlich} already in 1971. In the same
year Journal of Mathematical Sociology started and published Schelling's model 
of urban segregation \cite{schelling}, which is a modification of the Ising 
magnet at zero temperature. 1982 saw the start of two other lines of research 
by physicists on socio-economic questions \cite{galam,roehner}. 

Languages have been simulated on computers by decades, while the interest of 
physicists is more recent \cite{gomes,zanette}, triggered mostly by a model
of language competition \cite{abrams}.

We do not mention chemists since at present they play no major role in this 
field. However, the 1921 chemistry Nobel laureate F. Soddy \cite{soddy}, to 
whom we owe the "isotope" 
concept, already worked on economic, social and political theories, and 
his finance work of the 1930's was still cited in 2007. The present authors
try it the other way around: First apply physics to social sciences, and then
get the Nobel prize (for literature: science fiction.) 

\section{Some Models and Concepts}

Physicist Albert Einstein said that models should be as simple as possible, 
but not simpler. In this spirit we now introduce some basic physics models 
and concepts for readers from social sciences. They don't have to study
physics for many years, the following examples give the spirit. All models
are complex in the sense that the behaviour of large systems cannot be 
predicted from the properties of the single element.

\subsection{Cellular Automata}

Mathematicians denote cellular automata often as "interacting particle systems",
but since many other models or methods in physics use interacting particles,
we do not use this term here. A large $d$-dimensional 
lattice of $L^d$ sites carries variables
$S_i \; (i=1,2,\dots L^d)$ which can be either zero or one; more generally,
they are small integers between 1 and $Q > 2$. The lattice may be square (four
nearest neighbours), triangular (six nearest neighbours), or simple cubic
(also six nearest neighbours, but in $d=3$ dimensions); many other choices are
also possible. Time $t = 1,2, ...$ increases in steps. At each time step, 
each $S_i(t+1)$ is calculated anew from a deterministic or probabilistic rule
involving the neighbouring $S_k(t)$ of the previous time step. This way of
updating is called "simultaneous" or "parallel"; one may also use sequential
updating where $S_i$ depends in the current values of $S_k$; then the order
of updating is important: random sequential, or regular like a typewriter.

An example is a biological infection process: Each site $i$ becomes permanently
infected, $S_i = 1$, if at least one of its nearest neighbours is already
infected. Computers handle that efficiently if each computer word of, say, 32 
bits stores 32 sites, and if then 32 possible infections are treated at once 
by bit-by-bit logical-OR operations\ \cite{jphysa}.

\subsection{Temperature}

We know temperature $T$ from the weather reports, but in physics it enters 
according to Boltzmann into the probability $$p \propto \exp(-E/k_BT) \eqno(1)$$
to observe some configuration with an energy $E$. Here $T$ is the temperature
measured in Kelvin (about 273 + the Celsius or centigrade temperature),
and $k_B$ the Boltzmann constant relating the scales of energy and temperature.
For simplicity we now set $k_B = 1$, i.e. we measure temperature and energy
in the same units. If $g$ different configurations have the same energy, then 
$S = \ln(g)$ is called the entropy, and the probability to observe this energy 
is $\propto g \exp(-E/T) = \exp(-F/T)$ with the "free energy" $F = E - TS$.

In a social application we may think of peer pressure or herding: If your
neighbours drink Pepsi Cola, they influence you to also drink Pepsi,
even though at present you drink Coca Cola. Thus let $E$ be the number 
of nearest neighbours drinking Pepsi Cola, minus the number of Coke drinking
nearest neighbours. The probability for you to switch then is given by the
energy difference and equal to exp$(-2E/T)$ 
(or 1 if $E < 0$) in the Metropolis algorithm, or $1/(1 + \exp(2E/T)$ in the
Glauber or Heat Bath algorithm. In both cases there is a tendency to decrease
$E$. In the limit $T = 0$ one never makes a change which increases $E$, while
for small positive $T$ one increases $E$ with a low but finite probability.
In the opposite limit of infinite temperature, the energy becomes unimportant
and all possible configurations become equally probable. Neither zero nor 
infinite temperature are usually realistic.

In this sense, decreasing the energy $E$ is the most simple or most plausible 
choice, and the temperature measures the willingness or ability to deviate 
from this simplest option, e.g. to withstand peer pressure. But temperature
also incorporates all those random accidents of life which influence us but
are not part of the social model. For example, it may happen that there is 
no Pepsi Cola available even though all your neighbours drink Pepsi and you
want to follow them. Investors have to make their financial choices under the
influence of their clients, whose life is shaped by births, marriages, 
deaths, or other personal events which are not included explicitly into 
a financial market model. These accidents are then simulated by a finite
temperature, entering the probability that one does not follow the usual rule.

The ability to withstand peer pressure and the randomness of personal lives 
are in principle two different things, and if one wants to include them both
one needs two different temperatures $T_1$ and $T_2$, which do not exist in
traditional physics \cite{odor}.

\subsection{Ising Model}

In the model published by Ernst Ising in 1925, the variables $S_i$ are not 
0 or 1, but $\pm 1$: 

$$E = - \sum_{i,k} S_iS_k - B \sum_i S_i \eqno(2)$$ 
and for $B=0$ this corresponds to the above Coke versus Pepsi example. The
first summation runs over all neighbour pairs, the second over all sites.
Thus if site $i$ considers changing its variable, the energy change is
$\pm \Delta E = 2(\sum_k S_k - B)$ and enters through exp$(-\Delta E/T)$ 
into the probabilities to flip $S_i$; now $k$ runs over the neighbours of $i$
only. (If instead of flipping one $S_i$ one wants to exchange two different
variables $S_i$ and $S_j$, moving $S_i$ into site $j$ and $S_j$ into site $i$,
then one has to calculate the energy changes for both sites $i$ and $j$ in this
"Kawasaki" kinetics.) A computer program and pictures from its application
are given elsewhere in this Encyclopedia \cite{stauffer}.

In physics, the $S_i$ are magnetic dipole moments of the atoms, often called 
spins, and $B$ is proportional to the magnetic field. Usually, physicists
write an exchange constant $J$ before the first sum, but we set $J = 1$ for
simplicity here. The model was invented to describe ferromagnetism, like
in the elements iron, cobalt or nickel. Later it was found to describe 
liquid-vapour equilibria and other phase transitions. We know that 
iron at room temperature is magnetic, and this corresponds to the fact that 
for $0 < T < T_c$ and zero field $B$ the Ising model has the majority of its 
spins in one direction (either mostly +1 or mostly --1), while for $T > T_c$
half of the spins point in one and the other half in the opposite direction.
The magnetisation $M = \sum_i S_i$, often normalised by the number $L^d$ of
spins, is therefore an order parameter. The critical temperature $T_c$ is
often named after Pierre Curie.

In one dimension, we have $T_c$ = 0; in the square lattice in two dimensions
we know $T_c = 2/\ln(1 + \sqrt 2)$ exactly, while on the simple cubic lattice
$T_c \simeq 4.5115$ is estimated only numerically. Of course, one has 
generalized the model to more than nearest neighbours, to more than two 
states $\pm 1$ for each spin, and to disordered lattices and networks.

\subsection{Percolation} 
Simpler than the Ising model but less useful is percolation theory, reviewed 
more thoroughly in this Encyclopedia in the section edited by M. Sahimi. Each
site of a large lattice is randomly occupied with probability $p$, empty with 
probability $1-p$, and clusters are sets of occupied neighbouring sites. 
There is one percolation threshold $p_c$ such that for 
$p< p_c$ only finite clusters exist, for $p > p_c$ also one infinite 
cluster, and at $p = p_c$ even several infinite clusters may co-exist, which
are fractal: The number of occupied sites belonging to the infinite
clusters varies at $p_c$ as $L^D$ where $D$ is the fractal dimension. Here
"infinite" means: spanning from one end of the sample of $L^d$ sites to the
opposite end, or: increasing in average number of sites with a positive power 
of $L$. In one dimension, again one has no phase transition ($p_c = 1$), on 
the square lattice $p_c \simeq 0.5927462$ and on the simple cubic lattice
$p_c \simeq 0.311608$ are known only numerically, with a fractal dimension
of 1, 91/48 and $\simeq 2.5$ in one to three dimensions.

In the resulting disordered lattices, each site has from 0 to $z$ neighbours, 
where $z$ is the number of neighbours in the ordered lattice $p = 1$. If one
neglects the possibility of cyclic links one finds $p_c = 1/(z-1)$ in this
Bethe lattice or Cayley tree. Near this percolation threshold the critical
exponents with which several quantities diverge or vanish are the same as
in the random graphs of Erd\"os and R\'enyi. But this percolation theory 
was published nearly two decades earlier, in 1941, by the later chemistry 
Nobel laureate P. Flory.
   
\subsection{Mean Field Approximations}

What is called "mean field" is called "representative agent" theory in 
economics, and is widespread in chemistry where the changes in the 
concentrations of various reacting compounds are approximated as functions
of these time-dependent concentrations. A particularly simple example is 
Verhulst's logistic equation $dx/dt = ax(1-x)$ , known as Bass diffusion
in economics. We now explain why this approximation is unreliable.

Let us return to the above Ising model of Eq.(2) and replace the $S_k$ there
by it's average $<S_k> = m = M/L^d$ which is a real number between --1 and
+1 instead of being just --1 or +1; $m$ is the normalised magnetisation. Then
the total energy $E$ is approximated as the sum over single energies $E_i$:
$$ E = \sum_i E_i\; \quad E_i = (-\sum_k <S_k> - B)S_i = - B'S_i$$
with a mean magnetic field $B' = B + \sum_k <S_k> = B + mz$ where $z$ again 
is the number of lattice neighbours. The system now behaves as if each 
spin $S_i$ is in an effective field $B'$ influenced only by the average
magnetisation $m$ and no longer directly by its neighbours $S_k$. The two
possible orientations of $S_i$ have the energies $\pm B'$, giving an average
$$ m = <S_i> = {\rm tanh}(B'/T) = {\rm tanh}[(B + zm)/T] \eqno (3a)$$ 
and thus a self-consistency equation for $m$. Expanding the hyperbolic tangent 
into a Taylor series for small $m$ and $B$ we get
$$ B = (1 - z/T)m + m^3/3 + \dots \eqno (3b)$$
which gives a Curie temperature $T_c = z$, since for $T<T_c$ the magnetisation
is $m = \pm [3(z/T-1)]^{1/2} \propto (T_c-T)^{1/2}.$
Similar approximations for liquid-vapour equilibria lead to the Van der Waals 
equation of 1872, which may be regarded as the first quantitative theory of
a complex phenomenon. ($m$ there is the difference between the liquid and the 
vapour density.) Nowhere in Eqs(2,3a) have we put in that there is 
a phase transition to ferromagnetism; it just arises from the very simple
interaction energy $S_iS_k$ between neighbouring spins, and similarly the
formation of raindrops emerges from the interaction between the molecules of
water vapour. The water molecule is the same H$_2$O in the vapour, the liquid 
or the ice phase.

But this nice approximation contradicts the results mentioned above. For the
chain, square and simple cubic lattice it predicts a $T_c = z = 2$, 4 and 6
while the correct values are 0, 2.2, and 4.5. Particularly in one dimension
it predicts a phase transition at a positive $T_c$ while no such transition
is possible: $T_c=0$. This was the main result of Ernst Ising's thesis in 1925.
And even in three dimensions, where the difference in $T_c$ between 4.5 and 6 
is less
drastic, the above square-root law for $m$ is wrong, since $m$ varies 
for $T$ slightly below $T_c$ roughly as $(T_c-T)^{0.32}$. Thus mean field
theory, Van der Waals equation, and similar approximations averaging over many
particles are at best qualitatively correct. They become exact when 
each particle interacts equally with all other particles. 

Analogously for percolation, Flory's approximation of neglecting cyclic links
and the Erd\"os-R\'enyi random graphs lead to results corresponding to mean 
field approximations and should not be relied upon in one, two, or three 
dimensions with links between nearest neighbours only.

For cellular automata a particularly drastic failure of analogous mean field 
approximations (differential equations) was given by Shnerb et al \cite{shnerb}
for a biological problem. Even simpler, many cellular automata on the square
lattice lead to blinking pairs of next-nearest neighbours: at even times 
one site of the pair is 1 and the other is 0, while at odd times the first is
0 and the second is 1. Averaging over many sites destroys these local 
correlations which keep the blinking pair alive.

\section{Applications}

A thorough review of ``sociophysics'' was given recently by Castellano et al 
\cite{fortunatoRMP}, a long list of reference by Carbone et al \cite{carbone}. 
Some work of social scientists is reviewed by 
Davidsson and Verhagen in the section on agent-based simulations in 
sociology, while Troitzsch in this section reviews both social scientists
and physicists. His book with Gilbert \cite{gilbert} is, of course, more 
complete. Thus we merely sketch here some the areas covered in greater detail
in the other articles or in the cited literature.

\subsection{Elections}

\begin{figure}
\begin{center}
\includegraphics[scale=0.4]{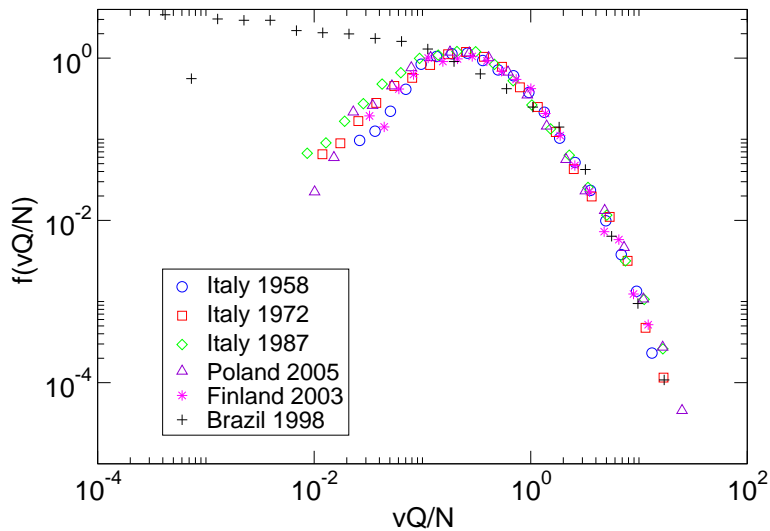}
\end{center}
\caption{The vote distribution in several countries and elections is a 
function only of the scaled variable $vQ/N$. From \cite{fortunato}.
}
\end{figure}

A social scientist may be interested to predict the fate of one particular
party or candidate in one particular election, or to explain it after this
election. A physicist, accustomed to electrons, hydrogen atoms and water
molecules being the same all over the world may be more interested to find
which universal properties all elections have in common. Figure 1, kindly sent
by Santo Fortunato, is an example. Let $v$ be the number of votes which a 
candidate got, $Q$ the number of candidates in that election, and $N$ the
total number of votes cast. Then the probability distribution $P(v,Q,N)$ for the
number of votes is actually a function $f(vQ/N)$ of only one scaled variable,
and that variable $vQ/N$ is the ratio of the actual number $v$ of votes to the
average number $N/Q$ of votes per candidate.  Various countries and 
various electsions, all using a proportional election system, gave the same 
curve $f(vQ/N)$ which is a parabola on this double-logarithmic plot and 
thus corresponds to a log-normal distribution. In Brazil, however, where 
the personality of a candidate play a major role, not only the party 
membership, the results were different. These authors \cite{fortunato} also
present a model to explain the log-normal distribution.

Other models for opinion dynamics are reviewed elsewhere in this Encyclopedia
\cite{stauffer}.

\subsection{Financial Markets}

\begin{figure}
\begin{center}
\includegraphics[scale=0.4,angle=-90]{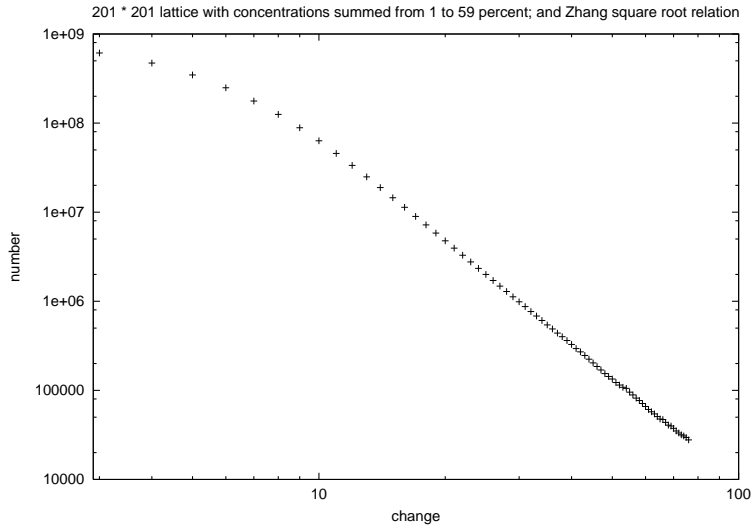}
\end{center}
\caption{Simulated return distribution in the Cont-Bouchaud percolation model
of stock markets \cite{cont}. The asymptotic slope to the right is about $-2.9$.
}
\end{figure}

Agent-based simulation of stock markets \cite{levy} are a typical example of
complex systems applications: In these models not the single agent but their
(unconscious) cooperation produces the ups and downs on the stock market, the
bubbles and the crashes. These models deal with the more or less random 
fluctuations, not with well founded market changes due to new inventions or 
major natural catastrophes. 

Real markets give at each time interval a return $r$ which is the relative 
change of the price. Typically, an index of the whole market like Dow Jones
changes each trading day by about one percent. Much larger fluctuations are
more rare, and the probability to have a change larger than $r$ decays for 
large $r$ as $1/r^3$: Fat tails. The sign of the change is barely predictable,
but its absolute value is: Volatility clustering. Thus in calm times when $|r|$
was small, tomorrow's $|r|$ probably is also small, whereas for turbulent times
with high $|r|$ in the past one should also expect a large $|r|$ tomorrow.
The daily weather behaves similarly: presumably tomorrow will be like today.
Perhaps even multifractality exists in real markets, similar to hydrodynamic
turbulence.

A simple model, going back to Bachelier more than a century ago, would throw 
a coin to determine whether the market tomorrow will go up or down. This simple
random-walk or diffusion model was shown by Mandelbrot in the 1960's not to 
describe a real market; it lacks fat tails and volatility clustering but may 
be good for monthly changes. Many better agent-based models have been invented 
during the last decade and reproduce these real properties, Fig. 2; the 
Cont-Bouchaud model is based on the above 
percolation theory \cite{cont}, while the Minority Game tells you it is better 
not to be with the big crowd \cite{challet}.

\subsection{Languages}

The versatility of human languages distinguishes us from the simpler 
communication systems of other living beings. With computers or mathematically
exact solutions \cite{komarova} models have been studied for the learning of a 
language by children or for the evolution of human languages out of simpler
forms. 

Closer to simulations in biology with the Darwinian selection of the fittest
are the models of competition between various languages of adult humans: Will
the Welsh language survive against English in Great Britain? Similar to
Lotka-Volterra equations for prey and predator in biology, some nonlinear
differential equations \cite{abrams} seem to describe the extinction of the
weaker language. Better statistics are available for the size distribution of 
languages, where "size" is the number of people speaking this language. Here
one model of de Oliveira et al found good agreement with reality, Fig.3; other 
models \cite{langssw} were less successful, inspite of many simulations from 
physicists. 

\begin{figure}
\begin{center}
\includegraphics[scale=0.8,angle=-90]{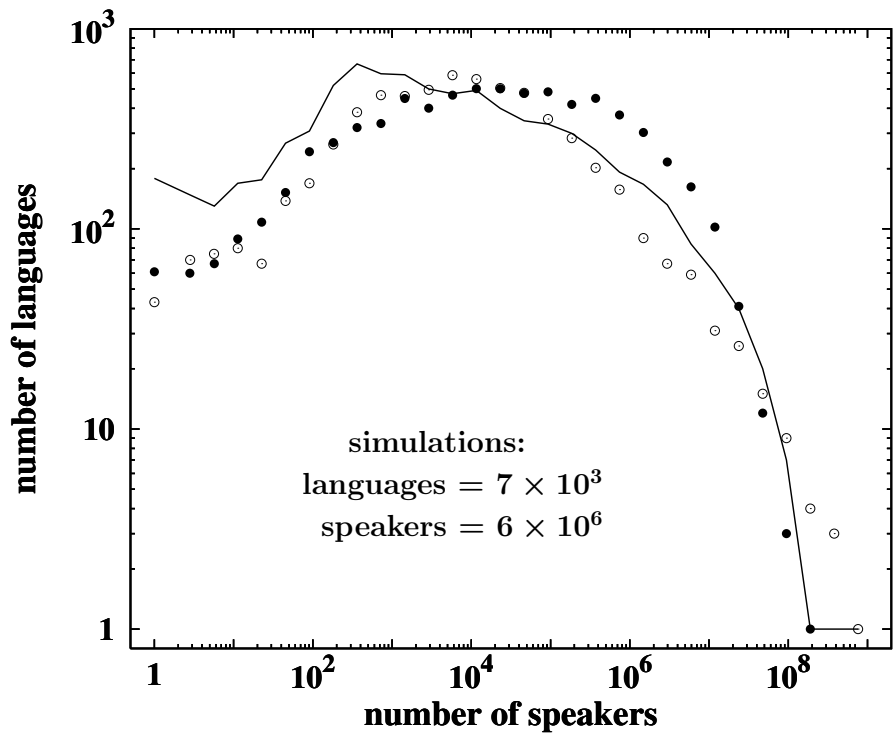}
\end{center}
\caption{Simulated size distributions for human languages (full circles, and 
line), compared with reality (open circles). From \cite{oliveira}.
}
\end{figure}

\section{Future Directions} 

The future should see more work in what we have shown here through our three 
figures: Searching for universal properties, or the lack of them, in the 
multitudes of models and in reality. Biology became a real science when the
various living beings were classified into horses, mammals, vertebrates
etc. Within each such taxonomic set all animals have certain things in common,
which animals in other taxonomic sets do not share. This check for universality
is different from improving our ability to ride horses. Thus making money on 
the stock market, or explaining the crash of 1987, is nice, but investigating
the exponents of the fat tails, Fig.2,  of all markets may give us more insight 
into what drives a market and what differences exist between different markets. 
Winning one particular election and predicting the winner is important, but 
universal scaling properties as in Fig.1 may help us to understand democracy
better. Preventing the extinction of French language in Canada is important
for the people there, but explaining the overall statistics of languages in
Fig.3 is relevant globally.

It is in these general aspects where the methods of mathematics and physics
seem to be most fruitful. One specific problem is better solved by the local
people who know that problem best, not by general simplified models.
A useful future approach for interacting agents would be their realisation
by neural network models \cite{wischmann}.

\end{document}